\documentclass[11pt]{article}
\begin{document}
\title
{G\"{o}del black hole, closed timelike horizon and the study of particle emissions}
\author
{ Sourav Bhattacharya,\footnote{sbhatt@bose.res.in}\\
S. N. Bose National Centre for Basic Sciences, \\
JD Block, Sector III, Salt Lake, Kolkata -700098, India,\\
and \\
 Anirban Saha, \footnote{anirban@iucaa.ernet.in}\\
Department of Physics, West Bengal State University,\\
Barasat, North 24 Paraganas,\\ West Bengal, India.\\
}

\maketitle
\abstract

 {We show that a particle, with positive orbital angular momentum, following an outgoing null/timelike geodesic, shall never reach the closed timelike horizon (CTH) present in the $(4+1)$-dimensional rotating G\"{o}del black hole space-time. Therefore a large part of this space-time remains inaccessible to a large class of geodesic observers, depending on the conserved quantities associated with them. We discuss how this fact and the existence of the closed timelike curves present in the asymptotic region make the quantum field theoretic study of the Hawking radiation, where the asymptotic observer states are a pre-requisite, unclear. However, the semiclassical approach provides an alternative to verify the Smarr formula derived recently for the rotating G\"{o}del black hole. We present a systematic analysis of particle emissions, specifically for scalars, charged Dirac spinors and vectors, from this black hole via the semiclassical complex path method.}

\hskip 1cm

{\bf Keywords:}{ G\"{o}del, Hawking radiation, Closed timelike horizon
\section{Introduction}
The G\"{o}del universe \cite{Godel:1949} is a cosmological solution of Einstein's equations in $4$-dimensions with pressure less dust satisfying weak energy condition, and a negative cosmological constant
\begin{eqnarray}
ds^2= -dt^2 + d\rho^2 + \alpha^{2}\left(\sinh^{4}\rho-\sinh^{2}\rho \right) d\psi^2+ dz^2 - 2\sqrt{2} \alpha~\sinh^{2}\rho~ dt d\psi ,
\label{god49}
\end{eqnarray}
where $\rho$ is a dimensionless radial variable, $\psi$ is the coordinate along a Killing field whose orbits are closed curves. $\alpha$ is a parameter determined by the energy density of the dust and the cosmological constant. The space-time (\ref{god49}) is spatially homogeneous but unlike the FRW cosmological solutions, it has a rotation parametrized by $\alpha$ \cite{Godel:1949}. In fact, a universe with rotation was the principal motivation to this solution. As can be seen in the metric (\ref{god49}), there exist naked and closed timelike curves (CTCs) when $g_{\psi \psi}\leq~ 0$. On the other hand, since this space-time is spatially homogeneous, at each point of the space-time there is a CTC \cite{Godel:1949}. For discussions on various geometric properties of (\ref{god49}) see \cite{Reboucas1}-\cite{Reboucas2}.

A few years ago a solution of Einstein's equations without any cosmological constant in the $(4+1)$-dimensional minimal supergravity was found \cite{Gauntlett:2003}. The bosonic part of the matter consists of a $U(1)$ Chern-Simons gauge field in $5$-dimensions. The metric and the  $U(1)$ gauge $1$-form $(A_a)$ are  
\begin{eqnarray}
ds^2&=&-dt^2+dr^2 +\frac{r^2}{4}d \theta^2 +\left(\frac{r^2}{4}-j^2 r^4 \cos^2 \theta\right)d \phi^2 +\left( \frac{r^2}{4}-j^2 r^4 \right)d \psi^2 \nonumber \\
&& + 2\left(\frac{r^2}{4}\cos\theta-j^2 r^4 \right)\cos\theta d\phi d\psi - 2 j r^{2}dt d \psi - 2 j r^{2}\cos \theta dt d \phi, 
\label{m1} \\
A_{a}&=&\frac{\sqrt{3}}{2}jr^2 \left[ (d \psi)_a+\cos{\theta}(d \phi)_a \right],
\label{m1a}
\end{eqnarray}
where $(\theta,~\phi,~\psi)$ are polar angles with $\psi$ being periodic, and $j$ is a parameter. As shown in \cite{Gauntlett:2003}, this solution mimics the salient features of the $4$-dimensional  G\"{o}del universe (\ref{god49}). The solution (\ref{m1}) is homogeneous and has a rotation parametrized by $j$. Moreover it is evident from  Eq. (\ref{m1}) that the Killing field $(\partial_{\psi})^a$, whose orbits are closed curves, is timelike for $r > \frac{1}{2 j}$. Since the solution (\ref{m1}) is homogeneous, we get an infinite and continuous sequence of CTCs throughout the space-time as in the $4$-dimensional G\"{o}del space-time. 

 An extreme Reissner-N\"{o}rdstrom black hole solution embedded in the $5$-dimensional G\"{o}del universe was found in \cite{Herdeiro:2003}. 
A solution, with the same matter content and without requiring to preserve any supersymmetry, representing a rotating black hole embedded in the $5$-dimensional G\"{o}del universe (\ref{m1}) was immediately found \cite{Hashimoto:2003} 
\begin{eqnarray*}
ds^2 & = & -u(r)dt^2-2g(r)\left( \cos \theta~ d \phi dt+  d \psi dt \right)+\frac{r^4}{\Delta(r)} dr^2+
 2 \left(h(r)+\frac{r^2}{4}\right)\cos \theta d \phi d \psi \nonumber\\ 
&& + \frac{r^2}{4}~d\theta^2 + \left(h(r) \cos^2 \theta + \frac{r^2}{4}\right)d \phi^2+\left(h(r) + \frac{r^2}{4}\right)d \psi^2, 
\end{eqnarray*}
where
\begin{eqnarray}
u(r)&=&1-\frac{2 M}{r^2},\qquad 
g(r)=jr^2+\frac{M a}{r^2},\qquad 
h(r)=-j^2 r^2(r^2+2 M)+\frac{M a^2}{2 r^2} ,\nonumber\\
\Delta(r)&=& r^4-2 Mr^2+ 8 j M(a+2j M)r^2+2 M a^2.
\label{metricgodel}
\end{eqnarray}
 The gauge $1$-form $A_a $ is given by Eq. (\ref{m1a}). The solution (\ref{metricgodel}) has two spherically symmetric horizons $\left(r_{{\rm{H}}},~r_{{\rm{in}}}\right)$~ defined by $\Delta=0$. Also it is evident from the metric that $g_{\psi \psi}=0$ at some  $r=r_{\rm{C}}$, i.e., the orbits of $(\partial_{\psi})^a$ are closed timelike curves for $r>r_{\rm{C}}$. We shall call $r_{\rm C}$ as closed time like horizon (CTH); also known as velocity of light surface (VLS) \cite{Gibbons}. $M$ is the mass parameter of the black hole whereas $a$ characterizes its angular momentum. When $M=0=a$, the metric (\ref{metricgodel}) reduces to the $5$-dimensional G\"{o}del universe (\ref{m1}). When $j=0$ it reduces to a $(4+1)$-dimensional Kerr black hole with two equal rotation parameters. We shall call space-time (\ref{metricgodel}) as Kerr-G\"{o}del black hole. A charged generalization of this solution can be seen in \cite{Herdeiro2}. As can be seen from the metric, the Kerr-G\"{o}del black hole (\ref{metricgodel}) asymptotically reaches the $5$-dimensional G\"{o}del space-time (\ref{m1}). 

To avoid any breakdown of causality by the presence of the CTCs at least for finite $r$, and also recalling the results in connection to the stability analysis \cite{Abdalla:2005}, we shall assume weakly rotating G\"{o}del background, i.e., ${\cal{O}}(j^2)$ terms will be ignored for finite $r$ in our analysis. It can be seen in Eq. (\ref{metricgodel}) that by doing this one pushes the CTCs off to large distances.

The Kerr-G\"{o}del black hole has two horizons $(r_{\rm{H}},~r_{\rm{in}})$ defined by $\Delta=0$. In the small $j$ approximation they take the form
\begin{eqnarray}
r_{{\rm{H}}}& = & \left\{M(1-4aj) + \sqrt{M^{2} \left( 1 - 8 a j \right)- 2 M a^{2}}\right\}^{\frac{1}{2}}, \label{H} \nonumber\\
r_{{\rm{in}}}& = & \left\{M(1-4aj) - \sqrt{M^{2} \left( 1 - 8 a j \right)- 2 M a^{2}}\right\}^{\frac{1}{2}}.
\label{C}
\end{eqnarray} 
Here $r_{\rm{H}}$ is the black hole event horizon, and $r_{\rm{in}}<r_{\rm{H}}$ is the inner horizon. We will be concerned only with $r_{{\rm{H}}}$. The surface gravity ($\kappa$) of the event horizon is 
\begin{eqnarray}
\kappa =\frac{r_{ {\rm{H}}}^2 - r_{{\rm{in}}}^2} { r_{ {\rm{H}}} \sqrt{  r_{ {\rm{H}}}^4+2Ma^2}}.
\label{kappa}
\end{eqnarray} 
The space-time also has an ergosphere defined by $g_{tt}=0$, i.e., at $r=\sqrt{2M}$.  

The G\"{o}del kind of solutions are interesting in the context of string theory. In \cite{Herdeiro:2003} the $5$-dimensional extreme Reissner-N\"{o}rdstrom-G\"{o}del black hole was uplifted to type IIB supergravity. Later  applying $T$-duality, the pure G\"{o}del solution (\ref{m1a}) was shown to be related to $pp$-waves \cite{Boyda}. For discussions of various exotic properties of G\"{o}del kind of solutions in the context of gauged supergravity and string theory see e.g. \cite{Hashimoto:2003}, \cite{Herdeiro3} -\cite{Wu}. In \cite{Abdalla:2005} the massless scalar field perturbation of a Schwarzschild-G\"{o}del black hole ($a=0$, in Eq. (\ref{metricgodel})) was studied showing that stable quasinormal modes exist only in the small $j$ limit. The parameter space of the Kerr-G\"{o}del black hole was analyzed numerically in \cite{Mann2:2007}. In \cite{Chen:2008, Li} the gray-body factor for the effective potential for a massless scalar field was estimated for the Kerr-G\"{o}del black hole. 

Since the Kerr-G\"{o}del space-time (\ref{metricgodel}) is not asymptotically flat, the computation of conserved charges and their variations are more subtle than the usual. This issue was addressed in \cite{Barnich:2005} where the resulting Smarr formula was shown to be in full agreement with the first law of black hole thermodynamics. Therefore it would be highly interesting
to get further insight into Hawking radiation \cite{Hawking1, Traschen} for various particle species in Kerr-G\"{o}del space-time. However, the motion of the emitted particle in this space-time should be much different than that in asymptotically flat ones. It is not difficult to anticipate that the most non-trivial feature of this motion will be related to the presence of the closed timelike horizon (CTH). In this paper we show that the outgoing particles, at least those having positive angular momentum, following a null/time-like geodesic, shall never reach the CTH and hence a large part of this space-time remains inaccessible to a large class of geodesic observers. The degree of inaccessibility depends on the conserved quantities corresponding to various Killing fields, associated with these observers.  Consequently, the notion of asymptotic observers becomes unclear in Kerr-G\"{o}del space-time. This makes the construction of a quantum field theory of the Hawking radiation, where the asymptotic observer states are a pre-requisite, problematic. This feature of the G\"{o}del space-time motivates us to apply the semiclassical treatment. We present a systematic analysis of the Hawking radiation, specifically for scalars, charged Dirac spinors and vectors, from rotating G\"{o}del black hole via this semiclassical method.

The semiclassical tunneling method is an alternative approach to model particle creation by black holes \cite{Kraus:1994}-\cite{Med}. This method has been successfully applied to $s$-wave scalar emissions for a wide class of black holes e.g. \cite{Quing} - \cite{Mann4}, to spinor emissions e.g. \cite{ mannspin}-\cite{Quan3} \footnote{For an exhaustive review and list of references on tunneling see e.g. \cite{rb}.}. For a comprehensive analysis of scalar emission in rotating
black hole background see \cite{Ang}. For a general analysis including backreaction see \cite{Med}. 


 The basic scheme of the semiclassical tunneling method is to compute the imaginary part of the `particle' action which gives the emission probability from the event horizon. From this emission probability one identifies the temperature of the radiation. The earliest works in this context can be found in \cite{Kraus:1994, Kraus:1996}. Following these works, an approach, called the null geodesic method, was developed \cite{Wilczek:2000, Parikh2:2004}. There exists, also, another way to model black hole evaporation via tunneling called complex path analysis \cite{Paddy1:1999, Paddy2:2001, Paddy3:2002} which we wish to apply here. This method involves writing down, in the semi-classical limit `$\hbar \to 0$', a Hamilton-Jacobi equation from the matter equation(s) of motion, treating the horizon as a singularity in the complex plane (which is a simple pole for all known solutions) and then complex integrating the equation across that singularity to obtain an imaginary contribution. This method would be quantitatively outlined in due courses in the following sections. 

The paper is organized as follows. In the next section we consider the motion of an outgoing particle following a geodesic. In Sect.-$3$ we compute, using complex path analysis, the zeroth order semiclassical tunneling probability and Hawking temperature for a scalar particle for the Kerr-G\"{o}del black hole. In Sect.-$4$ we present a suitable basis expansion of the $\gamma$-matrices in that space-time and compute the tunneling probability and the temperature for a charged spin up/down Dirac spinor. We also give an outline of the computation with a general $4$-component spinor. Here we show that similar results hold for vectors as well. We conclude in Sect.-$5$ with a brief discussion of our results and suggestions  of future directions of studies. We shall ignore the back reaction of the test matter. We shall work in a unit in which $c=1=G$, but will retain $\hbar$ throughout.

\section{Timelike/null geodesics and the CTH}
In this section we shall study the motion of an outgoing particle following a geodesic.
We shall confine our attention, for simplicity, to the equatorial `plane' $(\theta=\frac{\pi}{2})$. Tangent to this plane, let $v^a$ be the particle's $4$-velocity. We know that if $\chi^{a}$ is a Killing field and $v^{a}$ is a vector tangent to  a geodesic, i.e., $v^{a} \nabla_{a}v^{b} = 0 $, the quantity $v^{a}\chi_{a}$ is conserved along that geodesic. Thus, we may define the conserved quantities of motion $L$, $E$ and $L_{\phi}$ with respect to the Killing fields $\left( \partial_{\psi}\right)^{a}$, $\left( \partial_{t}\right)^{a}$ and $\left( \partial_{\phi}\right)^{a}$ respectively for our present space-time by
\begin{eqnarray}
L=\left( \partial_{\psi}\right)_{a}v^{a}=\left(h(r)+\frac{r^2}{4} \right)\dot{\psi}-g(r)\dot{t},\quad E=-\left( \partial_{t}\right)_{a}v^{a}=u(r)\dot{t}+g(r)\dot{\psi}, \quad L_{\phi} = \left( \partial_{\phi}\right)_{a}v^{a}=\frac{r^{2}}{4} \dot{\phi},
\label{geo1}
\end{eqnarray}
where the metric functions $(u,~g,~h)$ are defined in Eq. (\ref{metricgodel}) and the `dot' denotes derivative with respect to the proper time (an affine parameter) for a timelike (null) geodesic. 

Now, for a timelike/null particle following a geodesic we have
\begin{eqnarray}
-u(r)\dot{t}^2-2g(r)\dot{t}\dot{\psi}+\left(h(r)+\frac{r^2}{4} \right)\dot{\psi}^2+ \frac{r^{2}}{4} \dot{\phi}^{2}   +\frac{r^4}{\Delta}\dot{r}^2 =-k,
\label{geo2}
\end{eqnarray}
where $k=1,~0$ for timelike and null particles respectively. Using Eq. (\ref{geo1}), we eliminate $\dot{t}$, $\dot{\phi}$ and $\dot{\psi}$ from Eq. (\ref{geo2}) to get
\begin{eqnarray}
-u\frac{\left[E\left(h+\frac{r^2}{4} \right)-Lg\right]^2} {\left[g^2 +u \left(h+\frac{r^2}{4} \right) \right]^2} &-&   2g \frac{ \left[ E\left(h+\frac{r^2}{4}\right) -Lg \right] \left[Eg+Lu\right]}   { \left[g^2+u \left(h+\frac{r^2}{4} \right) \right]^2} \nonumber\\
& + & \left(h+\frac{r^2}{4} \right)
\frac{\left(Eg+Lu\right)^2}{\left[g^2+ u \left(h+\frac{r^2}{4} \right) \right]^2}+  \frac{4 L_{\phi}
^{2}}{r^{2}} +\frac{r^4}{\Delta}\dot{r}^2=-k.\nonumber\\
\label{geo3} 
\end{eqnarray}
The CTH is defined by $g_{\psi \psi}=\left(h(r)+\frac{r^2}{4} \right)= 0$. Taking $\left(h(r)+\frac{r^2}{4} \right)\to 0$ in Eq. (\ref{geo3}) we obtain
\begin{eqnarray}
\dot{r}^2\Big\vert_{{\rm CTH}}=-\frac{\Delta}{r^4}\left(\frac{2LE}{g}+\frac{L^2u}{g^2}+ \frac{4 L_{\phi} ^{2}}{r^{2}}+ k\right). 
\label{geo4} 
\end{eqnarray}
Let us now see the consequences of Eq. (\ref{geo4}). Since $E$ and $L$ are conserved along the geodesic they would not change their sign anywhere on the geodesic. $E$, being energy, is a positive definite quantity. Therefore the quantity within parenthesis in Eq. (\ref{geo4}) may be negative only when $L < 0$. On the other hand for $L \geq 0$, $\dot{r}$ becomes imaginary. Thus a geodesic with $L \geq 0 $ can never reach the CTH whereas those with $L<0$ may reach the CTH when the first term of Eq. (\ref{geo4}) is large enough to suppress the other terms. A particle with positive angular momentum should stop moving radially outwards when $\dot{r}=0$. Clearly, how far it can move depends on the value of the conserved quantities. Note that, our common intuition dictates that larger the energy of an outgoing particle, further it can move away from the black hole. But for the present space-time, it is clear from Eq. (\ref{geo4}) that a particle with positive angular momentum will never reach the vicinity of CTH however large its energy is.  
This phenomenon is independent of the small $j$ approximation and is also valid for the pure G\"{o}del space-time (\ref{m1}). 

The above analysis not only shows that the G\"{o}del space-times are very non trivial
and exotic but also
 provides an important insight about choosing the possible method of computing Hawking radiation. The quantum field theoretic study of Hawking radiation requires a well-defined asymptotic structure or the infinities \cite{Hawking1}. We have seen for Kerr--G\"{o}del space-time that only those geodesic observers having $L < 0$ may reach an asymptotic region (\ref{geo4}). Even if one manages to reach radial infinity with some negative value of $L$ the existence of CTCs there implies non-causal behavior of the space-time, therefore posing a major obstacle in following Hawking's quantum field theoretic treatment of particle creation \cite{Hawking1}. Hence we cannot make our study as an observer in that asymptotic region. Neither any point inside the CTH can be considered as infinity. Since different $L\geq 0$ observers have different turning points
(owing to the CTH) inside the CTH depending on their conserved quantities, the general mode of treatment should have an absolute local sense\footnote{An interesting point to note here is that, the de Sitter black hole space-times also do not have well-defined asymptotic structure due to the cosmological event horizon. But a quantum field theory of Hawking radiation can be done at least for eternal de Sitter black holes by treating each of the past and future horizons as past and future infinities \cite{Traschen}. When one considers the past black hole horizon as the ${\cal I}^{-}$, one considers the future cosmological horizon as ${\cal I}^{+}$ and vice-versa. But the Kerr--G\"{o}del space time do not have such cosmological horizons to be treated as infinities.}. It is this subtlety that motivates us to apply the semiclassical tunneling method to study Hawking radiation. This method only explicitly uses the near black hole horizon geometry. In the reminder of this paper we shall extensively study Hawking radiation of scalar, spinor and vector particles via tunneling.

\section{Scalar emission}
We begin by considering the massive Klein-Gordon equation
\begin{eqnarray}
\frac{1}{\sqrt{-g}}\partial_{a}\bigg(\sqrt{-g}g^{a b}\partial_{b}
\Phi(t,~r,~\theta,~\phi,~\psi)\bigg) +\frac{ \widetilde{m}^{2}}{\hbar^2} \Phi=0,
\label{WE}
\end{eqnarray}
where $g$ is the determinant of the metric. For the ansatz 
\begin{eqnarray}
\Phi\left(t,~r,~\theta,~\phi,~\psi\right) = A\left(t,~ r,~ ,\theta,~\phi,~ \psi\right) \exp^{\frac{i}{\hbar}I\left(t,~ r,~\theta, ~\phi, ~\psi\right)}, 
\label{scalaranshatz}
\end{eqnarray}
in the zeroth order of the semi-classical limit `$\hbar \to 0$', Eq. (\ref{WE}) takes the form
\begin{eqnarray}
A \left(g^{ab} \partial_{a}I \partial_{b}I  - \widetilde{m}^{2}\right) = 0.
\label{wkb1}
\end{eqnarray}
Since the space-time (\ref{metricgodel}) has three Killing fields $\{\partial_{t},~ \partial_{\phi},~ \partial_{\psi}\}$, we may make the following ansatz for the mode
\begin{eqnarray}
I = -Et + m\phi + \lambda \psi + U\left(r,~ \theta \right).
\label{I} 
\end{eqnarray}
 Substituting the ansatz (\ref{I}) in Eq. (\ref{wkb1}) we obtain the following Hamilton--Jacobi equation

\begin{eqnarray}
\left(\frac{\partial U\left(r,~ \theta \right)}{\partial r}\right)^{2} + \frac{4 r^{2}}{\Delta}\left( \frac{\partial U\left(r,~ \theta \right)}{\partial \theta} \right)^{2} & = &  \frac{r^{4}}{\Delta^{2}}\left[ \left(r^{4} + 2 M a^{2}\right)\left(E - \lambda \Omega \right)^{2} - \frac{4 \Delta \lambda^{2} r^{2} }{\left(r^{4} + 2 M a^{2}\right)} \right. \nonumber \\
&&\qquad \left. - \frac{4 \Delta \left(m - \lambda \cos \theta \right)^{2}}{r^{2} \sin^{2} \theta} - \Delta \widetilde{m}^{2} \right],
\label{hjs} 
\end{eqnarray} 
where the function
\begin{eqnarray}
\Omega \left( r \right)= \frac{g^{t\psi}}{ g^{tt}}=-\frac{g_{t \psi}}{g_{\psi \psi}} =\frac{4 \left(j r^{4} + M a\right)}{\left(r^{4} + 2 M a^{2}\right)}
\label{omega} 
\end{eqnarray}
can be interpreted as the coordinate angular velocity $\left(\frac{d \psi}{dt} \right)_{\phi=\rm{const.}}$, of a particle moving along the timelike vector field $\left (\partial_{t} - \frac{g_{t \psi}}{g_{\psi \psi}}\partial _{\psi}\right)^a $, which is  orthogonal to $\left(\partial_{\psi}\right)^a$. 

To compute the tunneling probability we have to compute the imaginary part of $U\left(r, ~\theta \right)$ \cite{Paddy1:1999}. We note that the right hand side of Eq. (\ref{hjs}) has singularities at the black hole event horizon $(r=r_{\rm{H}})$, i.e., when $\Delta = 0$, and/or at $\theta = 0,~ \pi$. Hence the imaginary contribution to $U\left(r, ~\theta \right)$ may come from both the singularities via complex integration. 

In the near horizon limit $\Delta \to 0$, the right hand side of Eq. (\ref{hjs}) becomes effectively spherically symmetric (since on the horizon $\Delta=0$ everywhere, even the axial singularities have no contribution). One possible way to make the left hand side correspond to this spherical symmetry is to assume $U\left(r, ~\theta\right) = V\left(r \right) + W \left(\theta\right)$ as $r \to r_{\rm{H}}$, so that the first term becomes manifestly spherically symmetric whereas the second term does not. Thus on the horizon we must put $\frac{\partial W \left( \theta \right)}{\partial \theta} = 0 $. Hence infinitesimally close to the horizon Eq. (\ref{hjs}) effectively becomes 
\begin{eqnarray}
\frac{\partial V \left(r \right)}{\partial r} & = & \pm  \frac{r^{2}}{\Delta} \sqrt{\left(r^{4} + 2 M a^{2}\right)}\left(E - \lambda \Omega \right)\Bigg \vert_{r \rightarrow ~r_{\rm{H}}},
\label{hjs1} 
\end{eqnarray} 
where the $+ ~(-)$ corresponds to the outgoing (ingoing) mode.

 Now we integrate Eq. (\ref{hjs1}) across the horizon using complex path analysis \cite{Paddy1:1999, Paddy2:2001, Paddy3:2002}. We lift the singularity to complex plane : $\left( r-r_{\rm{H}}\right) \rightarrow  \left( r-r_{\rm{H}} \pm i \epsilon\right)$~ and bypass the event horizon around an infinitesimal semi-circle. For both outgoing and incoming particle we choose anti-clockwise contours. After complex integrating Eq. (\ref{hjs1}) around $r_{\rm{H}}$ we find
\begin{eqnarray}
 V_{\pm} \left(r_{{\rm{H}}}\right) = \pm \pi i \left(E - \lambda \Omega_{H}\right) \frac{r_{\rm{H}} \sqrt{r_{\rm{H}}^{4} + 2 M a^{2}}}{r_{\rm{H}}^{2} - r_{\rm{in}}^{2}} = \pm \frac{i \pi r_{{\rm{H}}}{}^{2}\left(E - \lambda \Omega_{{\rm{H}}}  \right) }{\sqrt{2 \left( M - 4 M a j - 2 a^{2} - 8 a^{3} j \right)}},
\label{set9s}
\end{eqnarray}
where $\Omega_{{\rm{H}}}= \Omega({r_{{\rm{H}}}})$.
Recalling the sign convention for an outgoing or incoming particle, the ansatz made in Eq.s (\ref{scalaranshatz}) and (\ref{I}), and the fact that  $W(\theta)$ is trivial at the horizon, the emission (absorption) probabilities $P_{{\rm{E}}}~(P_{{\rm{A}}})$ across the horizon are expressed as
\begin{eqnarray}
P_{{\rm{E}}} \propto  \exp\left[ - \frac{2}{\hbar} {\rm {Im}} \left\{V_{+} ( {r_{{\rm{H}}}} ) \right\}\right], \qquad 
P_{{\rm{A}}} \propto  \exp\left[ - \frac{2}{\hbar} {\rm {Im}} \left\{V_{-}({r_{{\rm{H}}}})\right\}\right].
\label{p}
\end{eqnarray}
 It follows from Eq. $(\ref{p})$ after using Eq. (\ref{set9s}),
\begin{eqnarray}
\frac{P_{{\rm{E}}}}{P_{{\rm{A}}}} \propto 
\exp \left[-  \frac{\left(E - \lambda \Omega_{H}\right)}{ \frac{\hbar \left(r_{\rm{H}}^{2} - r_{\rm{in}}^{2}\right)} {4 \pi r_{\rm{H}} \sqrt{r_{\rm{H}}^{4} + 2 M a^{2}}}   }\right] = 
\exp\left[ - \frac{ \left(E - \lambda \Omega_{{\rm{H}}} \right)}{ \frac{\hbar \sqrt{2 \left( M - 4 M a j - 2 a^{2} - 8 a^{3} j \right)}}{4 \pi r_{{\rm{H}}}^{2}}} \right].
\label{D}
\end{eqnarray}
We note from Eq. (\ref{D}) that $ P_{{\rm{E}}} \ll P_{{\rm{A}}}$. This can be understood as the `smallness' of quantum effects. In fact by defining proper normalization one may take $P_{{\rm{A}}} \to 1$.

The factor $\left(E - \lambda \Omega_{{\rm{H}}}  \right)$ appearing in Eq. (\ref{D}) is  the eigenvalue of $i \hbar \left( \partial_t +\Omega_{{\rm{H}}} \partial _{\psi}\right)$. The Killing $1$-form $\left( \partial_t +\Omega_{{\rm{H}}} \partial _{\psi}\right)_a$ is future directed null at the horizon, and future directed timelike infinitesimally outside it.  Therefore $\left(E - \lambda \Omega_{{\rm{H}}} \right)$ can be interpreted as the energy of the particle as measured by an observer moving along that Killing field. $\left(E - \lambda \Omega_{{\rm{H}}}  \right) \geq ~0$ for a timelike or null emitted particle; whereas for an incoming particle, $\left(E - \lambda \Omega_{{\rm{H}}}  \right) < ~0$ corresponds to the superrradiant modes which we shall disregard. With the factor $\left(E - \lambda \Omega_{{\rm{H}}} \right)$ being regarded as the energy of the emitted particle, we now identify from Eq. (\ref{D}) the temperature of the emission or the Hawking temperature of the event horizon \cite{Paddy1:1999}, which after using Eq. (\ref{H}), takes the form 
\begin{eqnarray}
T_{{\rm{H}}} = \frac{\hbar \left(r_{\rm{H}}^{2} - r_{\rm{in}}^{2}\right)} {4 \pi r_{\rm{H}} \sqrt{r_{\rm{H}}^{4} + 2 M a^{2}}} 
=\frac{\hbar  \left( M - 4 M a j - 2 a^{2} - 8 a^{3} j \right)^{\frac{1}{2}}}{2\sqrt{2} \pi \left[M - 4Maj + \left\{ M^{2} \left(1 - 8aj\right) - 2Ma^{2}\right\}^{\frac{1}{2}}\right]}.
\label{set11}
\end{eqnarray} 
One immediately identifies from Eq. (\ref{kappa}) that $T_{\rm{H}}=\frac{\kappa\hbar}{4 \pi}$. Note that when we set $a = 0 =r_{\rm{in}} $ in Eq. (\ref{set11}) we recover the well known  result for a  Schwarzschild black hole : $T_{\rm{H}}=\frac{\hbar}{4 \pi r_{\rm{H}}}$ \cite{Hawking1}.

The above is in full agreement with the predicted thermodynamic behavior of the Kerr-G\"{o}del black hole \cite{Barnich:2005}. The $\theta$ independence of Eq.s (\ref{D}) and (\ref{set11}) can be understood as the manifestation of the spherical symmetry of the horizon. Calculation of this temperature by null geodesic method can be found in \cite{Mann2:2007}. $j=0$ in Eq. (\ref{set11}) recovers the temperature of a $5$-dimensional Kerr black hole, whereas $a=0$ recovers the result for a Schwarzschild-G\"{o}del black hole. Note that here we have not assumed the usual $s-$wave nature of the emission, but demonstrated explicitly how the $\theta-$part of the modes becomes trivial near the horizon due to the horizon's spherical symmetry. We shall see that this will be true for spinors and vectors as well.

\section{Spinor and vector emissions}
The Dirac equation for a massive spinor in a curved space-time (or in the flat space-time written in an arbitrary coordinate system) is given by 
\begin{equation}
i\gamma ^{a }D_{a }\Psi +\frac{\widetilde{m}}{\hbar}\Psi =0. 
\label{Dirac}
\end{equation}%
For a spinor with a charge $q$, minimally coupled to a gauge $1$-form $A_a$, the gauge covariant derivative operator $D_a$ is defined as
\begin{eqnarray}
D_{a } \Psi &=& \left(\partial _{a } - \Gamma_{a } -\frac{i~q}{\hbar}A_a \right) \Psi. 
\label{D1} 
\end{eqnarray}
Here $\partial_a$ are the ordinary derivatives and $\Gamma_a$ are the spin connection matrices
\begin{eqnarray}
\Gamma _{a } &=&\frac{1}{8}[\gamma ^{(\mu)},\gamma ^{(\nu) }]\omega_{a(\mu) (\nu)}, 
\label{D3}
\end{eqnarray}
where $\omega$ are the Ricci rotation coefficients. For some choice of orthonormal bases
 $e_{(\mu)}{}^{a}$,~ $\omega_{a(\mu) (\nu)}:= e_{(\mu)}{}^{b}\nabla_{a} e_{(\nu)b} $. For the G\"{o}del black hole we are considering, $A_a$ is given by Eq. (\ref{m1a}). Now and hereafter the Latin indices will represent space-time and the Greek indices in parenthesis will represent (local) Lorentz frame. So the $\gamma^{(\mu)}$ appearing in Eq. (\ref{D3}) correspond to the Minkowski space (which is $(4+1)$-dimensional in our case) satisfying the anticommutation relation 
\begin{eqnarray}
\{\gamma^{(\mu) },~\gamma^{(\nu)}\}=2\eta^{(\mu)( \nu) } {\bf{I} },
\label{Lgamma}
\end{eqnarray} 
where $\eta^{(\mu)( \nu)} = {\rm {diag}}\{-1,~ 1,~ 1,~ 1,~1\}$ is the $(4+1)$-dimensional Minkowskian metric and ${\bf{I} }$ is the identity matrix. On the other hand, the $\gamma ^{a }$ appearing in Eq. (\ref{Dirac}) can be expanded as $\gamma ^{a }=e_{(\mu)}{}^{a}\gamma^{(\mu)}$. Then, from the definition $\eta^{(\mu)(\nu) }e_{(\mu)}{}^{a}e_{(\nu)}{}^{b}:= g^{ab}$ and from Eq. (\ref{Lgamma}) we obtain a generalized anti-commutation relation
\begin{eqnarray}
\{\gamma ^{a },\gamma ^{b}\}=2g^{ab } \;{\bf{I} }. 
\label{algebra}
\end{eqnarray}

 To proceed further, therefore, we have to choose a suitable representation of $\gamma^a$ subject to Eq. (\ref{algebra}). We choose the following representation for our space-time (\ref{metricgodel}) 
\begin{eqnarray}
\gamma ^{t} &=&\sqrt{- g^{tt}} \gamma^{(0)}, \qquad
\gamma^{r}=\sqrt{g^{rr}}\gamma ^{(3)},\qquad \gamma ^{\theta }= \sqrt{g^{\theta \theta}}\gamma ^{(1)}, \nonumber\\
\gamma ^{\phi } &=& \left[g^{\phi \phi} - \frac{\left\{g^{\phi \psi}-\frac{g^{t \phi} g^{t\psi}}{g^{tt}}  \right\}^{2}}{\left\{g^
{\psi \psi} - \frac{\left(g^{t \psi}\right)^{2}}{g^{tt}}\right\}}-\frac{(g^{t \phi})^2}{g^{tt}}  \right]^{\frac{1}{2}}\gamma^{(4)} +   \frac{  \left\{ g^{\phi \psi} -\frac{g^{t \phi}g^{t \psi}}{g^{tt}}\right\}  }{\left\{g^{\psi \psi} - \frac{\left(g^{t \psi}
\right)^{2}}{g^{tt}}\right\}^{\frac{1}{2}}}\gamma^{(2)}- \frac{g^{t \phi}}{\sqrt{-g^{tt}}} \gamma^{(0)},\nonumber\\
\gamma^{\psi}&=& \left[g^{\psi \psi} - \frac{\left(g^{t \psi}\right)^{2}}{g^{tt}}\right]^{\frac{1}{2}}\gamma^{(2)} -  \frac{g^{t \psi}}{\sqrt{-g^{tt}}}\gamma^{(0)},
\label{curvedgammas}
\end{eqnarray}
where $\gamma^{(4)}$ is defined to be the fifth spacelike $\gamma$-matrix in the $(4+1)$-dimensional Minkowski space-time satisfying the algebra of Eq. (\ref{Lgamma}). 

An usual representation of $\gamma^{(\mu)}$ ($\mu = \{0,~1,~2,~3\}$) is
\begin{eqnarray}
\gamma ^{(0)} &=&\left( 
\begin{array}{cc}
0 & I \\ 
-I & 0
\end{array}
\right), \quad \gamma^{(1)}=\left( 
\begin{array}{cc}
0 & \sigma ^{1} \\ 
\sigma ^{1} & 0%
\end{array}%
\right),  \quad
\gamma ^{(2)} = \left( 
\begin{array}{cc}
0 & \sigma ^{2} \\ 
\sigma ^{2} & 0%
\end{array}%
\right), \quad  \gamma ^{(3)}=\left( 
\begin{array}{cc}
0 & \sigma ^{3} \\ 
\sigma ^{3} & 0%
\end{array}%
\right). \nonumber\\
\label{chiralgammas}
\end{eqnarray}%
The $\sigma$ are the Pauli spin matrices 
\begin{equation}
\sigma ^{1}=\left( 
\begin{array}{cc}
0 & 1 \\ 
1 & 0%
\end{array}%
\right), \quad  \sigma ^{2}=\left( 
\begin{array}{cc}
0 & -i \\ 
i & 0%
\end{array}%
\right), \quad \sigma ^{3}=\left( 
\begin{array}{cc}
1 & 0 \\ 
0 & -1%
\end{array}%
\right)  \label{paulis}.
\end{equation}
We have yet to choose a suitable representation of the fifth spacelike matrix $\gamma^ {(4)}$. We simply choose $ \gamma^ {(4)}=\gamma^5$ , i.e.,
\begin{eqnarray}
\gamma^{(4)} = i \gamma^{(0)} \gamma^{(1)}\gamma^{(2)}\gamma^{(3)}.
\label{4}
\end{eqnarray}
Now we are ready to look into Eq. (\ref{Dirac}).

The spin up  ansatz for the Dirac particle has the form 
\begin{eqnarray}
\psi (t,~r,~\theta,~\phi,~\psi) 
&=&\left[ 
\begin{array}{c}
A(t,~r,~\theta,~\phi,~\psi) \\ 
0 \\ 
B(t,~r,~\theta,~\phi,~\psi ) \\ 
0%
\end{array}%
\right] \exp \left[ \frac{i}{\hbar }I(t,~r,~\theta,~\phi,~\psi)\right]. 
\label{spin up}
\end{eqnarray}%
Inserting the ansatz (\ref{spin up}) into Eq. (\ref{Dirac}) and taking the semiclassical limit ~`$\hbar \rightarrow 0$' we get
\begin{eqnarray}
\left\{-\gamma^{a}  \partial_{a}I + q\left(\gamma^{\phi} A_{\phi} 
+ \gamma^{\psi} A_{\psi}\right) + \widetilde{m} \right\}\left[ 
\begin{array}{c}
A(t,~r,~\theta,~\phi,~\psi) \\ 
0 \\ 
B(t,~r,~\theta,~\phi,~\psi) \\ 
0%
\end{array}%
\right]  = 0, 
\label{wkb2}
\end{eqnarray}
where $A_{\phi}$ and $A_{\psi}$ are the components of the gauge $1$-form given in Eq. (\ref{m1a}). Now making the usual ansatz as before
\begin{eqnarray}
I = -Et + m \phi + \lambda \psi + U \left(r,~\theta \right),
\label{spinor_anshatz}
\end{eqnarray}
and using Eq.s (\ref{curvedgammas}), (\ref{chiralgammas}), (\ref{paulis}), (\ref{4}) we get from Eq. (\ref{wkb2}) the following set of four equations after neglecting ${\cal {O}}(j^2)$ terms 
\begin{eqnarray}
\sqrt{- g^{tt}} B\left( E - \frac{g^{t\psi}}{ g^{tt}}\lambda + \frac{g^{t\psi}}{ g^{tt}} q A_{\psi}\right) - \sqrt{g^{rr}} B \partial_{r} U  + \widetilde{m}A + \left\{g^{\phi \phi} - \frac{\left(g^{\phi \psi}\right)^{2}}{\left(g^{\psi \psi} - \frac{\left(g^{t \psi}\right)^{2}}{g^{tt}}\right)}\right\}^{\frac{1}{2}}\left(m - q A_{\phi} \right)A & = & 0, \nonumber\\
\label{set1}\\
\sqrt{- g^{tt}} A\left(E  - \frac{g^{t\psi}}{ g^{tt}}\lambda + \frac{g^{t\psi}}{ g^{tt}} q A_{\psi}\right) + \sqrt{g^{rr}} A \partial_{r} U - \widetilde{m}B + \left\{g^{\phi \phi} - \frac{\left(g^{\phi \psi}\right)^{2}}{\left(g^{\psi \psi} - \frac{\left(g^{t \psi}\right)^{2}}{g^{tt}}\right)}\right\}^{\frac{1}{2}}\left(m - q A_{\phi}\right) B  & = & 0, \nonumber\\
\label{set2}
\end{eqnarray}
\begin{eqnarray}
B \left[ \sqrt{ g^{\theta \theta}} \partial_{\theta} U + i\frac{ g^{\phi \psi}}{\left(g^{\psi \psi} - \frac{\left(g^{t \psi}\right)^{2}}{g^{tt}}\right)^{\frac{1}{2}}} \left(m - q A_{\phi}\right)  + i\left(g^{\psi \psi} - \frac{\left(g^{t \psi}\right)^{2}}{g^{tt}}\right)^{\frac{1}{2}} \left(\lambda  - q A_{\psi}\right)\right] & = & 0, 
\label{set3} \\
A \left[ \sqrt{ g^{\theta \theta}} \partial_{\theta} U + i\frac{ g^{\phi \psi}}{\left(g^{\psi \psi} - \frac{\left(g^{t \psi}\right)^{2}}{g^{tt}}\right)^{\frac{1}{2}}} \left(m - q A_{\phi}\right) + i \left(g^{\psi \psi} - \frac{\left(g^{t \psi}\right)^{2}}{g^{tt}}\right)^{\frac{1}{2}}\left(\lambda  - q A_{\psi}\right)\right] & = & 0. 
\label{set4}
\end{eqnarray}
 For nontrivial solutions of Eq. (\ref{Dirac}), $A$ and $B$ cannot be identically zero everywhere. Thus we may eliminate them from Eq.s (\ref{set1}) and (\ref{set2}) to obtain an expression for $\left( \frac{\partial U}{\partial r}\right)$.  
Next, from either of the Eq.s (\ref{set3}) and (\ref{set4}) we determine $\sqrt{g^{\theta \theta}} \partial_{\theta} U $, square it and add to $\left( \frac{\partial U}{\partial r}\right)^2$ to get the following Hamilton-Jacobi equation
\begin{eqnarray}
\left( \frac{\partial U}{\partial r}\right)^2 + g^{\theta \theta} \left(\frac{\partial U}{\partial \theta} \right)^2 = &-\frac{ g^{tt}}{g^{rr}}\left[\left(E - \lambda \Omega - q A_{\psi} \Omega \right)^{2}   - \frac{\widetilde{ m} {}^{2}}{g^{tt}}  + \frac{\left(m - q A_{\phi}\right)^{2}}{g^{tt}}\left\{g^{\phi \phi} - \frac{\left(g^{\phi \psi}\right)^{2}}{\left(g^{\psi \psi} - \frac{\left(g^{t \psi}\right)^{2}}{g^{tt}}\right)}\right\}\right]-\nonumber \\ 
&\left[\frac{ g^{\phi \psi}}{\left(g^{\psi \psi} - \frac{\left(g^{t \psi}\right)^{2}}{g^{tt}}\right)^{\frac{1}{2}}} \left(m - q A_{\phi}\right) +  \left(g^{\psi \psi} - \frac{\left(g^{t \psi}\right)^{2}}{g^{tt}}\right)^{\frac{1}{2}}\left(\lambda  - q A_{\psi}\right)\right]^2. 
\label{set5}
\end{eqnarray}
The function $\Omega(r)$ has the same interpretation as in the scalar case.

 To this end we again apply complex path analysis and follow the same procedure as described in the previous section. One can check that Eq. (\ref{set5}) has singularities at the horizon $(r_{\rm{H}})$, i.e., when $\Delta = 0$, and/or  at $\theta = 0, ~ \pi$. Let us first consider the horizon. The $\Delta = 0$ singularity makes the right hand side of Eq. (\ref{set5}) spherically symmetric and again we can write $U \left(r, ~ \theta \right) = V \left(r \right) + W \left( \theta \right)$ infinitesimally close to the horizon. Now using complex integration we find from Eq. (\ref{set5})
\begin{eqnarray}
 V_{\pm} \left(r_{{\rm{H}}}\right) =   \pm \frac{i \pi r_{{\rm{H}}}{}^{2}\left(E - \lambda \Omega_{{\rm{H}}} - q A_{\psi}\left(r_{\rm{H}}\right) \Omega_{{\rm{H}}} \right) }{\sqrt{2 \left( M - 4 M a j - 2 a^{2} - 8 a^{3} j \right)}},
\label{set9}
\end{eqnarray}
along with a trivial $W(\theta)\big \vert_{r=r_{\rm{H}}}$. The $+~(-)$ sign corresponds to the outgoing (incoming) particle and $\Omega_{{\rm{H}}}= \Omega({r_{{\rm{H}}}})$. 

 Thus recalling the ansatz made in Eq.s (\ref{spin up}) and (\ref{spinor_anshatz}), we see that the emission (absorption) probabilities $P_{{\rm{E}}}~(P_{{\rm{A}}})$ across the event horizon $(r_{\rm{H}})$ have the same form given in Eq. (\ref{p}).
So it follows using Eq.s (\ref{set9}) and (\ref{m1a}),
\begin{eqnarray}
\frac{P_{{\rm{E}}}}{P_{{\rm{A}}}} \propto \exp\left[ - \frac{\left(E - \lambda \Omega_{{\rm{H}}} - \frac{\sqrt{3}}{2} q j r_{\rm{H}}^{2}  \Omega_{{\rm{H}}}\right) }{\frac{\hbar \sqrt{2 \left( M - 4 M a j - 2 a^{2} - 8 a^{3} j \right)}}{4 \pi r_{{\rm{H}}}^{2}}} \right].
\label{Dspin}
\end{eqnarray}
Eq. (\ref{Dspin}) gives the Hawking temperature $(T_{\rm{H}})$ and it is identical to that of the scalar (Eq. \ref{set11}). Due to the spherical symmetry of the horizon the axial singularities $\theta=0,~\pi$ have no effect on the emission probability from the horizon. 

Similar analysis can also be performed for a spin-down particle giving the same Hawking temperature. However, for completeness, before we end this section we give an outline for computation with a general $4$-component wave function. We take the ansatz 
\begin{eqnarray}
\psi (t,~r,~\theta,~\phi,~\psi) 
&=&\left[ 
\begin{array}{c}
A e^{\frac{i}{\hbar}I_{\downarrow}}\\ 
C e^{\frac{i}{\hbar}I_{\uparrow}}\\ 
B e^{\frac{i}{\hbar}I_{\downarrow}}\\ 
D e^{\frac{i}{\hbar}I_{\uparrow}}
\end{array}
\right](t,~r,~\theta,~\phi,~\psi),
\label{spin gen}
\end{eqnarray}
where the $(\uparrow \downarrow)$ sign refers to spin up and spin down particles respectively. We know that any difference in the energy eigenvalues of spin up and spin down states should be ${\cal O}(\hbar)$ and hence this difference can safely be ignored in our semiclassical theory. Thus we may take 
\begin{eqnarray}
I_{\uparrow \downarrow}(t,~r,~\theta,~\phi,~\psi) = -Et + m\phi + \lambda \psi + U_{\uparrow \downarrow}\left(r,~ \theta \right).
\label{Ig} 
\end{eqnarray}
We now substitute this ansatz in Eq. (\ref{Dirac}). Making the following definitions
\begin{eqnarray}
\mu&=&\left( I_{\uparrow} -I_{\downarrow} \right),~
\epsilon_{1}=\sqrt{- g^{tt}} \left( E - \frac{g^{t\psi}}{ g^{tt}}\lambda + \frac{g^{t\psi}}{ g^{tt}} q A_{\psi}\right),\\
\epsilon_{2\uparrow \downarrow}&=&\sqrt{g^{rr}}  \partial_{r} U_{\uparrow \downarrow},~ 
\epsilon_{3\pm}=\widetilde{m} \pm \left\{g^{\phi \phi} - \frac{\left(g^{\phi \psi}\right)^{2}}{\left(g^{\psi \psi} - \frac{\left(g^{t \psi}\right)^{2}}{g^{tt}}\right)}\right\}^{\frac{1}{2}}\left(m - q A_{\phi} \right),\nonumber  \\
\epsilon_{4\uparrow}&=&\left[- \sqrt{ g^{\theta \theta}} \partial_{\theta} U_{\uparrow} + i\frac{ g^{\phi \psi}}{\left(g^{\psi \psi} - \frac{\left(g^{t \psi}\right)^{2}}{g^{tt}}\right)^{\frac{1}{2}}} \left(m - q A_{\phi}\right) + i \left(g^{\psi \psi} - \frac{\left(g^{t \psi}\right)^{2}}{g^{tt}}\right)^{\frac{1}{2}}\left(\lambda  - q A_{\psi}\right)\right],\nonumber  \\
\epsilon_{4\downarrow}&=&\left[\sqrt{ g^{\theta \theta}} \partial_{\theta} U_{\downarrow} + i\frac{ g^{\phi \psi}}{\left(g^{\psi \psi} - \frac{\left(g^{t \psi}\right)^{2}}{g^{tt}}\right)^{\frac{1}{2}}} \left(m - q A_{\phi}\right) + i \left(g^{\psi \psi} - \frac{\left(g^{t \psi}\right)^{2}}{g^{tt}}\right)^{\frac{1}{2}}\left(\lambda  - q A_{\psi}\right)\right],\nonumber \\
\label{abbr} 
\end{eqnarray}
we obtain the following set of four equations
\begin{eqnarray}
B\left(\epsilon_{1}- \epsilon_{2\downarrow}\right)+A \epsilon_{3+}+D e^{\frac{i \mu}{\hbar}}\epsilon_{4\uparrow}=0,~D\left(\epsilon_{1}+ \epsilon_{2\uparrow}\right)+C \epsilon_{3-}-B e^{-\frac{i \mu}{\hbar}}\epsilon_{4\downarrow}=0,\nonumber\\
A\left(\epsilon_{1}+ \epsilon_{2\downarrow}\right)+B \epsilon_{3-}+C e^{\frac{i \mu}{\hbar}}\epsilon_{4\uparrow}=0,~C\left(\epsilon_{1}- \epsilon_{2\uparrow}\right)+D \epsilon_{3+}+A e^{-\frac{i \mu}{\hbar}}\epsilon_{4\downarrow}=0.
\label{4eqs} 
\end{eqnarray}
Therefore, unlike our previous study with only spin up or spin down ansatz (Eq.s (\ref{set1})-(\ref{set4})), now the $r$  and $\theta$ derivatives do not give separate equations.
Eliminating the functions $(A,~B,~C,~D)$ from Eq.s (\ref{4eqs}) we obtain
\begin{eqnarray}
\left(\epsilon_{1}^2-\epsilon_{2\uparrow}^2- \epsilon_{3-}^2\right) \left(\epsilon_{1}^2-\epsilon_{2\downarrow}^2+ \epsilon_{3+}^2\right)+ \left(\epsilon_{1}-\epsilon_{2\downarrow}\right)\left(\epsilon_{1}+\epsilon_{2\uparrow}\right) \epsilon_{4\uparrow}\epsilon_{4\downarrow}
+  \left(\epsilon_{1}+\epsilon_{2\downarrow}\right)\left(\epsilon_{1}-\epsilon_{2\uparrow}\right) \epsilon_{4\uparrow}\epsilon_{4\downarrow}+\epsilon_{4\uparrow}^2\epsilon_{4\downarrow}^2=0.\nonumber \\
\label{resultingeq} 
\end{eqnarray}
We now take the near horizon limit $(\Delta \to 0)$ in Eq. (\ref{resultingeq}). As before we can break $U_{\uparrow \downarrow}(r,~\theta)=V_{\uparrow \downarrow}(r)+W_{\uparrow \downarrow}(\theta)$ and apply the same arguments to obtain
\begin{eqnarray}
\left[\left\{ \left(  \frac{\epsilon_{1}}{\sqrt{g^{rr}}}  +\partial_{r} V_{\downarrow}\right) \left( \frac{\epsilon_{1}}{\sqrt{g^{rr}}} -\partial_{r} V_{\downarrow} \right)\right\}  \left\{ \left( \frac{\epsilon_{1}} {\sqrt{g^{rr}}}+  \partial_{r} V_{\uparrow}\right) \left( \frac{ \epsilon_{1}}{ \sqrt{g^{rr}}} - \partial_{r} V_{\uparrow}\right)\right\}\right]_{\Delta \to 0} =0,
\label{factored} 
\end{eqnarray}
along with a trivial $W_{\uparrow \downarrow}(\theta)\big \vert_{r=r_{\rm{H}}}$. Thus the desired results follow.

Now we give an outline for computation for the vectors. We start with the following equation of motion
\begin{eqnarray}
\nabla_{a}F^{ab} = m^{2} A^{b}.
\label{vector1}
\end{eqnarray}
Eq. (\ref{vector1}) can be written as 
\begin{eqnarray}
\nabla_{a}\nabla^{a} A_{b} - R_{b a} A^{a} + \nabla_{b}
\left(\nabla_{c}A^{c}\right) = m^{2} A_{b}.
\label{vector2}
\end{eqnarray}
 Expanding $A_{b}$ in the orthonormal basis i.e., $A _{a }=e^{(\mu)}{}_{a}A_{(\mu)}$ and using the usual semiclassical ansatz for each $A_{(\mu)} =g_{\mu} \exp^{\frac{i}{\hbar}I(x)}$ and the identity $\nabla_{a}A^{a} = 0$, one can immediately see that Eq. (\ref{vector2}) reduces to four Klein--Gordon equations in the zeroth order of the semiclassical theory and hence the result for scalar or spinor is reproduced.

\section{{Discussions}}
In this work we have discussed the motion of a particle following a null/time-like geodesic in the Kerr-G\"{o}del black hole space-time and shown that at least the particles having positive orbital angular momentum $L$ shall never reach the closed time-like horizon (CTH). So it is evident that the notion of asymptotic observers is not clear within the CTH. Again, beyond the CTH there exists a continuous sequence of closed time-like curves (CTC) and hence the causal structure of space-time is lost. These issues pose high obstacles in formulating a field theory of Hawking radiation in Kerr-G\"{o}del space-time. However, in \cite{Barnich:2005} a Smarr formula was derived for the Kerr-G\"{o}del black hole. This demands an extensive analysis of particle emission in this context. To perform this task we have done the semiclassical treatment for scalars, charged Dirac spinors and vectors. The expressions of the emission probability and hence the Hawking temperature do not contain any parameter (mass, charge, spin) of the matter and same for all particle species. This is due to the fact that back reaction of the matter is ignored in our treatment and also because the near horizon limit was taken. The emission probability and the Hawking temperature turns out to be in full agreement with \cite{Barnich:2005}. This, indeed, verifies the thermodynamic nature of the Smarr formula.

 Another important thing to note is the effect of the rotation of the background (parametrized by $j$) on emissions. For simplicity we take $a \to 0$ in Eq. (\ref{set11}) to get
\begin{eqnarray}
T_{\rm{H}}=\frac{\hbar}{4 \pi \sqrt{2M}} \left(1+2aj\right) +{\cal{O}}(j^2).
\label{a0}
\end{eqnarray}
This shows an increment in the Hawking temperature with $j$. Thus the scalar or neutral spinor emission probabilities (given by Eq. (\ref{D})) also increase with $j$. This can naively be interpreted as the `centrifugal' effect on the particles due to the rotation of the G\"{o}del background. 

We have considered the geodesics in the equatorial plane only. It would be very interesting to study them for the entire space-time. To do that the idea of Killing tensor as in the $4$-dimensional Kerr black hole may be useful. Since a considerable number of the geodesics cannot be arbitrarily extended due to the CTH,
it may be possible that an outgoing emitted particle following a geodesic with sufficient kinetic energy may come back and enter the black hole. Note that this effect will be there
along with the usual grey-body effect.
This suggests that the evaporation rate for the G\"{o}del black holes may be quite different than the rotating black holes in asymptotically flat space-times. However this dynamics requires further study. Also, the construction of a satisfactory quantum field theory of the Hawking radiation in this space-time remains as an interesting problem. 


\section*{Acknowledgment}
S. Bhattacharya sincerely acknowledges Amitabha Lahiri for useful discussions and for going through the manuscript.
\vskip 1cm


\begin{thebibliography}{99}
\bibitem{Godel:1949}
 K.~G\"{o}del,
 Rev.\ Mod.\ Phys. \ {\bf 21}, 447 (1949).
 


\bibitem{Reboucas1}
M.~J.~Reboucas and J.~Tiomno, 
Phys.\ Rev.\ D {\bf 28}, 1251 (1983). 



\bibitem{Reboucas2}
M.~J.~Reboucas {\it et al}, 
Phys.\ Rev.\ D {\bf 32}, 3309 (1985).






\bibitem{Gauntlett:2003}
J.~P.~Gauntlett {\it et al},
 Class.\ Quant.\ Grav. \ {\bf 20}, 4587 (2003).
 


\bibitem{Herdeiro:2003}
C.~A.~R.~ Herdeiro,
 Nucl.\ Phys.\ B \ {\bf 665}, 189 (2003).




\bibitem{Hashimoto:2003}
 E.~Gimon and A.~Hashimoto,
 Phys.\ Rev.\ Lett. \ {\bf 91}, 021601 (2003).


\bibitem{Gibbons}
 G.~W.~Gibbons  {\it et al},
[arXiv: hep-th/0504080].
\bibitem{Herdeiro2}
C.~A.~R.~Herdeiro, 
Class. \ Quant. \ Grav. {\bf 20}, 4891 (2003). 


\bibitem{Abdalla:2005}
R.~A.~Konoplya and E.~Abdalla,
 Phys.\ Rev.\ D \ {\bf 71}, 084015 (2005).


\bibitem{Boyda}
E.~K.~Boyda {\it et al}, 
Phys.\ Rev.\ D\ {\bf 67}, 106003 (2003).

\bibitem{Herdeiro3}
C.~A.~R.~Herdeiro {\it et al}, 
Phys.\ Rev.\ D {\bf 69}, 066010 (2004). 



\bibitem{Mann2:2005}
 R.~B.~Mann {\it et al}, 
JHEP  {\bf 0504}, 49 (2005).


\bibitem{Mann3:2005}
 R.~B.~Mann {\it et al}, 
 Phys.\  Lett.\ B  \ {\bf 620}, 1 (2005).











\bibitem{Wu}
S.~Q.~Wu,
Phys.\ Rev. \ Lett.\  {\bf 100}, 121301 (2008).



\bibitem{Mann2:2007}
 R.~Kerner and R.~B.~Mann,
 Phys. \ Rev. \ D {\bf 75}, 084022 (2007).                               



\bibitem{Chen:2008}
 S.~Chen {\it et al},
 Phys.\ Rev.\ D \ {\bf 78}, 064030 (2008).



\bibitem{Li}
 W.~Li {\it et al},                                         
Class.\ Quant. \ Grav.\ {\bf 26}, 055008 (2009). 



\bibitem{Barnich:2005}
G.~Barnich and G.~Compere,
 Phys.\ Rev.\ Lett. \ {\bf 95}, 031302 (2005).



\bibitem{Hawking1}
S.~W.~Hawking,
 Commun.\ Math.\ Phys. \ {\bf 43}, 199 (1975).
 

\bibitem{Traschen} J.~Traschen, An Introduction to Black Hole Evaporation {\it Mathematical Methods of Physics}, proceedings of the 1999 Londrina Winter School, editors A. Bytsenko and F. Williams, World Scientific (2000), arXiv:gr-qc/0010055.





\bibitem{Kraus:1994}
 P.~Kraus and F.~Wilczek,
[arxiv:gr-qc/9406042], Nucl. \ Phys.\ B \ {\bf 437}, 231 (1995).






\bibitem{Kraus:1996}
 P.~Kraus and E.~Keski-Vakkuri,
 Nucl. \ Phys.\ B \ {\bf 491}, 249 (1997).  



\bibitem{Wilczek:2000}
 M.~K.~Parikh and F.~Wilczek,
 Phys. \ Rev. \ Lett. {\bf 85}, 5042 (2000).



\bibitem{Parikh2:2004}  
 M.~K.~Parikh,
 Int.\ J. \ Mod. \ Phys.\  {\bf D13}, 2351 (2004).



\bibitem{Paddy1:1999}
 K.~Srinivasan and T.~Padmanabhan,
 Phys. \ Rev. \ D {\bf 60}, 24007 (1999).



\bibitem{Paddy2:2001}
 T.~Padmanabhan {\it et al}, 
 Mod. \ Phys. \ Lett. {\bf A 16}, 571 (2001).
 


\bibitem{Paddy3:2002}
 T.~Padmanabhan {\it et al},  
 Class. \ Quant. \ Grav. {\bf 19}, 2671 (2002).




\bibitem{Medved}
A.~J.~M.~Medved, 
Phys.\ Rev.\ D {\bf 66}, 12409 (2002). 




\bibitem{Quing}
Q.~Jiang {\it et al}, 
Phys.\ Rev.\  D\ {\bf 73},  064003 (2006).



\bibitem{zhang}
J.~Zhang and Z.~Zhao, 
Phys.\ Lett.\  B\ {\bf 638}, 110 (2006).        



\bibitem{Zhao} 
L.~Zhao,
Commun. \ Theor. \ Phys.\  {\bf 47}, 835, (2007).



\bibitem{Mann4}
R.~Kerner and R.~B.~Mann, 
Phys.\ Rev.\  D \ {\bf73}, 104010 (2006).




\bibitem{Iso}
F.~Wilczek {\it et al},
Phys.\ Rev.\ D\ {\bf 74}, 044017 (2006).



\bibitem{mannspin}
 R.~Kerner and R.~B.~Mann, 
Class.\ Quant.\ Grav. {\bf 25}, 095014 (2008). 


\bibitem{mannspin2}
 R.~Kerner and R.~B.~Mann,
Phys. \ Lett. \ B {\bf 665}, 277 (2008).



\bibitem{vanzo}
R.~Di~Criscienzo and L.~Vanzo, 
Europhys.\ Lett. \  {\bf 82}, 60001 (2008).



\bibitem{Li1:2008}
 R.~Li and J.~R.~Ren,  
Phys.\ Lett. \ B \  {\bf 661}, 370 (2008).


\bibitem{Li2:2008}
 R.~Li and J.~R.~Ren,
 Class.\ Quant.\ Grav.\ {\bf 25}, 125016 (2008).


\bibitem{Quan2}
 Q.~Q.~Jiang,
 Phys. \ Lett.\  B {\bf 666},  517 (2008).


\bibitem{Quan3}
 Q.~Q.~Jiang,
 Phys.\  Rev. \ D {\bf 78}, 044009 (2008).



\bibitem{rb} 
R.~Banerjee and B.~R.~Majhi,
 JHEP \ {\bf 0806}, 095 (2008). 


\bibitem{majhi1}
B.~R.~Majhi,
Phys.\ Rev.\ D {\bf 79}, 044005 (2009). 



\bibitem{majhi2}
B.~R.~Majhi and S.~Samanta,
[arXiv: hep-th/0901.2258].

\bibitem{Ang}
M.~Angheben {\it et al},
  JHEP \ {\bf 05}, 014 (2005).

\bibitem{Med}
 A.~Medved and E.~Vagenas, 
Mod. \ Phys. \ Lett. \ A {\bf 20}, 2449 (2005).

 
\end{thebibliography}
\end{document}